\documentclass{PoS}
\usepackage{amsmath}

\title{Physical matrix elements for $\Delta I = 3/2$ channel $K \to \pi \pi$ decays }

\ShortTitle{Physical matrix elements for $\Delta I = 3/2$ channel $K \to \pi \pi$
decays }

\author{\speaker{Matthew Lightman} \\
        Department of Physics, Columbia University, New York, NY 10027, USA\\
        E-mail: \email{lightman@phys.columbia.edu}}

\author{RBC and UKQCD collaborations}

\abstract{$K\to\pi\pi$ matrix elements of the electroweak operator $Q_{(27,1)}^{\Delta I=3/2}$ are calculated on the RBC/UKQCD $32^3 \times 64$, $L_s=16$ lattices, using 2+1 dynamical flavors and domain wall fermions, with an inverse lattice spacing of $a^{-1}=2.42(4)\text{ GeV}$.  Data is interpolated or extrapolated to energy conserving kinematics and a preliminary calculation of the experimental parameter $|A_2|$ is performed.}

\FullConference{The XXVI International Symposium on Lattice Field Theory \\
                 July 14 - 19, 2008\\
                 Williamsburg, Virginia, USA}

\begin{document}

\section{Introduction}
\vspace{-0.14 in}
There is much interest in precision lattice calculations of $K \to \pi \pi$ decays because they can yield information on the origin of the $\Delta I=\frac{1}{2}$ rule and CP violation in the Standard Model \cite{CPPACS,RBC}.  While many quenched calculations of these decays have been performed (for example \cite{RBC,Changhoan_thesis,Changhoan1,Changhoan2,Takeshi}), a full dynamical calculation has yet to be done.  Steps have been made here toward a complete calculation of this type.  In order to obtain reasonable precision we use 2+1 flavors of domain wall fermions (DWF) on a $24^3 \times 64$ or even $32^3\times 64$, $L_s=16$ lattice.


In physical $K\to\pi\pi$ decays, the kinematics are such that the pions have non-zero momentum in the CM frame.  Giving the pions momentum on the lattice can introduce a lot of noise due to the fact that one is projecting onto an excited state of the two pion system rather than the ground state. For this reason only pions with (nearly) zero momentum are simulated in this paper, and the introduction of significant pion momentum is left to future work.
\vspace{-0.14 in}
\section{Four Quark Operators and the Effective Hamiltonian}
\vspace{-0.14 in}
The weak interactions and the effects of the heavier quarks can be included in the lattice QCD simulation by evaluating matrix elements of an effective Hamiltonian \cite{Ciuchini,Buchalla}.  In particular we use the conventions of \cite{RBC}
\begin{equation}\label{H_eff}
\mathcal{H}_{\Delta S = 1}=\frac{G_F}{\sqrt{2}}V_{ud}V_{us}^*\sum\limits_{i=1}^{10} [z_i(\mu)+\tau y_i(\mu)]Q_i
\end{equation}
where $V_{kl}$ are CKM matrix elements, $z_i$ and $y_i$ are Wilson coefficients, $\tau=-\lambda_t/\lambda_u$ with $\lambda_j\equiv V_{jd}V_{js}^*$, and $\{Q_i,i=1,...,10\}$ are four quark operators.  Therefore we are interested in calculating matrix elements of the four quark operators $Q_i$ between a $K$ and a $\pi\pi$ state.  These operators can be split into $\Delta I=3/2$ and $\Delta I=1/2$ parts, where $\Delta I$ is the change in isospin induced by the operator.  They can then be further classified by how they transform under the chiral $SU(3)_L\times SU(3)_R$ symmetry, and the representations (27,1), (8,8), and (8,1) are all found among various of the operators \cite{CPPACS,RBC}.
\vspace{-0.14 in}
\section{Extraction of the Matrix Element on the Lattice}
\vspace{-0.14 in}
We calculate only matrix elements of the operator $Q_{(27,1)}^{\Delta I=3/2}$, the single operator that transforms as (27,1) under $SU(3)_L\times SU(3)_R$ and $\Delta I=3/2$ under isospin.  To simplify matters we calculate the unphysical matrix element $\langle \pi^+\pi^+|{Q'}_{(27,1)}^{\Delta I=3/2}|K^+\rangle$ which can be related to the physical matrix element $\langle \pi^+\pi^0|Q_{(27,1)}^{\Delta I=3/2}|K^+\rangle$ by the Wigner Eckhart theorem \cite{Changhoan_thesis} if ${Q'}_{(27,1)}^{\Delta I=3/2}$ is given by
\begin{equation}
{Q'}_{(27,1)}^{\Delta I=3/2}=\bar{s}\gamma_\mu(1-\gamma^5)d\bar{u}\gamma^\mu(1-\gamma^5)d
\end{equation}

To extract the matrix element of this operator we calculate the following correlation functions
\begin{equation}\label{correlators}
C_K(t,t_K)=\langle O_K(t)O^\dag_K(t_K)\rangle,\hspace{0.1 in} C_{\pi\pi}(t,t_{\pi})=\langle O_{\pi\pi}(t)O^\dag_{\pi\pi}(t_\pi)\rangle,\hspace{0.1 in} C_{\mathcal{O}}(t_{\pi},t,t_K)=\langle O_{\pi\pi}(t_{\pi})\mathcal{O}_W(t)O^\dag_K(t_{K})\rangle
\end{equation}
where $t_K < t < t_\pi$ and where the interpolating operators are given by
\begin{equation}\label{interp_op}
O^\dag_{K}(t)=\sum\limits_{{\bf x},{\bf y}}\bar{u}({\bf x},t)\gamma^5 s({\bf y},t),\quad O_{\pi\pi}(t)=\sum\limits_{{\bf x},{\bf y},{\bf x'},{\bf y'}}\bar{d}({\bf x},t)\gamma^5 u({\bf y},t)\bar{d}({\bf x'},t)\gamma^5 u({\bf y'},t) 
\end{equation}
\begin{equation}
\mathcal{O}_W(t)=\sum\limits_{{\bf x}}\bar{s}({\bf x},t)\gamma_\mu(1-\gamma^5)d({\bf x},t)\bar{u}({\bf x},t)\gamma^\mu(1-\gamma^5)d({\bf x},t)
\end{equation}
Note that the interpolating operators in (\ref{interp_op}) are such that the correlators use wall sources and wall sinks, and project onto the zero momentum kaon and nearly zero momentum two pion states.

For $t_K\ll t\ll t_{\pi}$ we expect the following quotient of correlators to show a plateau in $t$:
\begin{equation}\label{quotient}
\frac{C_{\mathcal{O}}(t_\pi,t,t_K)}{C_K(t,t_K) C_{\pi\pi}(t,t_\pi)}\sim\frac{\mathcal{M}}{Z^*_{\pi\pi}Z_K},\quad t_K\ll t\ll t_{\pi}
\end{equation}
Here $\mathcal{M}$ is the matrix element we wish to extract. $Z_K$ and $Z_{\pi\pi}$ appear in the normalization factors for the kaon and two pion correlators respectively, and it is only possible to extract $|\mathcal{M}|$.
\vspace{-0.14 in}
\section{Details of the Lattice Calculation}
\vspace{-0.14 in}
Calculations were carried out on the RBC/UKQCD $32^3\times 64$, $L_s=16$ 2+1 flavor domain wall fermion lattices.  The inverse lattice spacing for these lattices is 2.42(4) GeV \cite{Enno_proceedings}, corresponding to a physical volume of $(2.6\text{ fm})^3$.  The sea strange quark mass is always 0.03 in lattice units, and for the sea light quark mass (henceforth denoted by just $m_{sea}$) there is an ensemble with $m_{sea}=0.004$ and an ensemble with $m_{sea}=0.008$, both with 129 configurations.  For each ensemble, inversions are performed with the following valence masses: $m_{val}=$0.002, 0.004, 0.006, 0.008, 0.025, and 0.030.  We calculate matrix elements for all possible valence mass combinations such that $m_s\ge m_l$ where $m_s$ is the valence strange quark mass and $m_l$ is the valence light quark mass.

We add and subtract propagators with periodic and antiperiodic boundary condtions in order to double the effective time length and suppress around the world contributions.  The resultant periodic plus antiperiodic (P+A) propagator has a source at t=0 and the resultant periodic minus antiperiodic (P-A) propagator effectively has a source at t=64.  These provide the left and right walls for the kaon at $t_K=0$ and the two pions at $t_\pi=64$ respectively, and the time $t$ at which the operator is located is varied.  Twelve wall source propagators using a unit source distributed over a single time slice, each with a specific spin and color (fixed in Coulomb gauge), are computed on each configuration and used to evaluate the correlation functions in (\ref{correlators}).
\vspace{-0.14 in}
\section{Results}
\begin{figure}[htp]
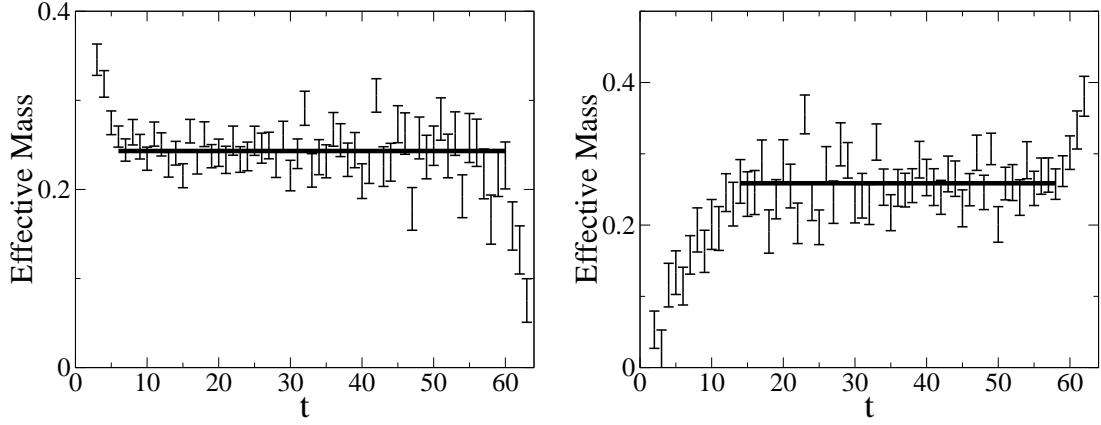

\centering
\includegraphics{kaon_meff}
\hspace{0.1 in}
\includegraphics{pipi_meff}
\hfill
\vspace{-0.1 in}
\caption{Effective mass plots and fitted values of the mass/energy for the kaon and two pion correlators with $m_{sea}=0.004$, $m_s=0.03$, and $m_l=0.004$.  \underline{Left:}  Kaon correlator.  \underline{Right:}  Two pion correlator.}\label{fig:kaon_pipi_meff}
\end{figure}

\begin{figure}[htp]
\centering
\includegraphics{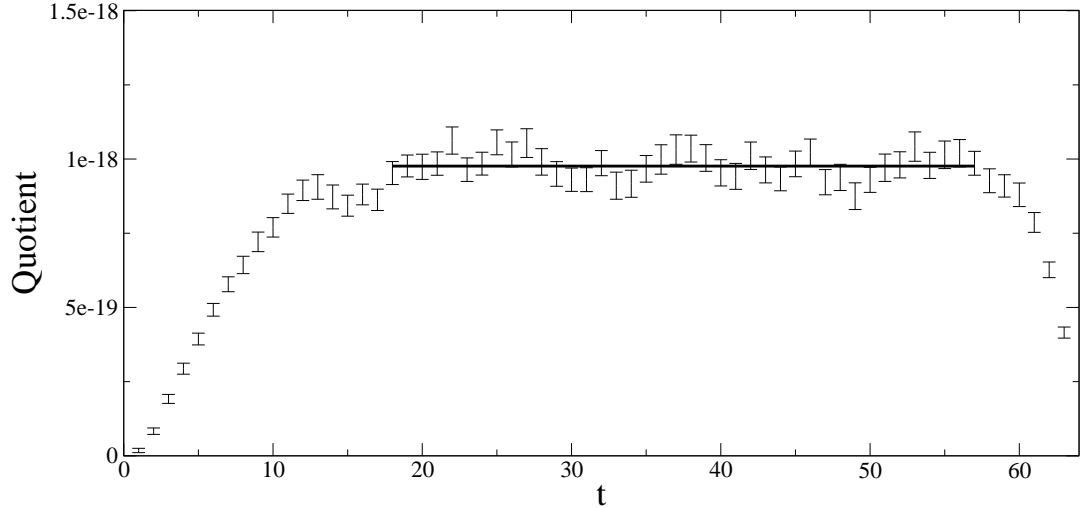}
\hfill
\vspace{-0.1 in}
\caption{Plot of the $K\to\pi\pi$ correlator divided by the kaon and two pion correlators for $m_{sea}=0.004$, $m_s=0.03$, and $m_l=0.004$.  The fit to the plateau is shown.}\label{fig:quotient_plot}
\end{figure}

\vspace{-0.14 in}
Effective mass plots of the kaon and two pion correlators are shown in Figure \ref{fig:kaon_pipi_meff}, for $m_{sea}=0.004$, $m_s=0.03$, and $m_l=0.004$.  When the $K\to\pi\pi$ correlator is divided by the kaon and two pion correlators as in (\ref{quotient}), we find that the quotient shows a plateau as expected.  The quotient and a fit to the plateau are shown for $m_{sea}=0.004$, $m_s=0.03$, and $m_l=0.004$ in Figure \ref{fig:quotient_plot}.

From the lattice matrix element $|\mathcal{M}|$ we wish to find the physical quantity $|A_2|$, as defined in \cite{RBC}, which can be compared to experiment.  $|\mathcal{M}|$ must be multiplied by the following factors in order to obtain $|A_2|$
\begin{enumerate}
\item The Wilson coefficient plus factors that appear in the effective Hamiltonian (\ref{H_eff}) multiplying the operator.  Wilson coefficients evaluated at $\mu=2$ GeV are interpolated from \cite{Sam_thesis}.
\item Numerical factors due to the Wigner Eckhart transformation that relates $K^+\to\pi^+\pi^0$ to $K^+\to\pi^+\pi^+$.
\item A lattice to $\overline{\rm MS}$ renormalization factor $Z_{B_K}^{\overline{MS}}(\mu)$ for the operator. It is evaluated at the same scale $\mu$ as the Wilson coefficient and is obtained from \cite{CKelly}.
\item A factor proportional to $\sqrt{m_KE_{\pi\pi}^2L^3}$, where $L=32a$ is the spatial extent of the lattice.  This performs a naive transformation between finite and infinite volume normalization, ignoring the effects of $\pi\pi$ scattering that have been analyzed in \cite{Lellouch_Luscher}.  We hope to include the more accurate correction of \cite{Lellouch_Luscher} in future work.
\end{enumerate}

Technically the contribution of the other two $\Delta I=3/2$ four quark operators to $|A_2|$ should also be included.  However, these have been found to be much smaller than the contribution of the (27,1) operator \cite{Changhoan_thesis}, so we neglect them.

A table of results for all of the different mass combinations can be found in Table 1.  The data for $|A_2|$ is new and was not presented at the time of the talk.  Note that none of the mass combinations correspond to a process in which energy is conserved, although $m_{sea}=0.004$, $m_s=0.03$, $m_l=0.004$ comes close.  Notice also that changing $m_{sea}$ doesn't have much of an effect.  This insensitivity to the sea quark mass justifies doing a linear interpolation/extrapolation to the unitary point $m_{sea}=m_l$ with respect to the sea quark mass for a given $m_l$.  The sea {\it strange} quark mass, however, is always fixed at 0.03, so we are stuck with a non-unitary strange quark if $m_s\ne 0.03$.

\begin{table}[htp]
\label{tb:data}
\caption{Kaon masses, two pion energies, matrix element values, and $|A_2|$ results for different valence mass combinations. Each line of data for $m_{sea}=0.004$ has the corresponding line of data for $m_{sea}=0.008$ below it so that the insensitivity of the data to sea quark mass can be seen.}
\small
\begin{center}
\begin{tabular}{|c|c|c|c|c|c|c|}
\hline
$m_l$&$m_s$&$m_{sea}$&$m_K$ (MeV)&$E_{\pi\pi}$ (MeV)&$|\mathcal{M}|$&$|A_2|$ ($10^{-8}$ GeV)\\
\hline
0.002&0.002&0.004&237.1(9)&485(2)&0.001166(27)&1.510(36)\\
0.002&0.002&0.008&241.4(9)&493(3)&0.001148(29)&1.521(41)\\
0.002&0.004&0.004&274.9(9)&485(2)&0.001189(28)&1.657(40)\\
0.002&0.004&0.008&279.0(8)&493(3)&0.001159(28)&1.651(42)\\
0.002&0.006&0.004&308.1(9)&485(2)&0.001213(30)&1.790(45)\\
0.002&0.006&0.008&312.1(8)&493(3)&0.001178(28)&1.774(45)\\
0.002&0.008&0.004&338.1(9)&485(2)&0.001236(31)&1.910(50)\\
0.002&0.008&0.008&341.9(8)&493(3)&0.001200(29)&1.893(48)\\
0.002&0.025&0.004&528(1)&485(2)&0.001399(50)&2.701(97)\\
0.002&0.025&0.008&531(1)&493(3)&0.001409(47)&2.770(94)\\
0.002&0.03&0.004&572(1)&485(2)&0.001438(55)&2.89(11)\\
0.002&0.03&0.008&576(1)&493(3)&0.001472(55)&3.01(12)\\
0.004&0.004&0.004&307.7(9)&626(2)&0.001459(29)&2.773(56)\\
0.004&0.004&0.008&311.7(8)&632(2)&0.001408(26)&2.721(51)\\
0.004&0.006&0.004&337.5(8)&626(2)&0.001466(30)&2.918(60)\\
0.004&0.006&0.008&341.3(8)&632(2)&0.001418(25)&2.869(53)\\
0.004&0.008&0.004&364.8(8)&626(2)&0.001477(30)&3.056(64)\\
0.004&0.008&0.008&368.6(8)&632(2)&0.001433(26)&3.012(55)\\
0.004&0.025&0.004&545(1)&626(2)&0.001609(41)&4.07(10)\\
0.004&0.025&0.008&549.0(9)&632(2)&0.001576(33)&4.044(87)\\
0.004&0.03&0.004&588(1)&626(2)&0.001644(44)&4.32(12)\\
0.004&0.03&0.008&592(1)&632(2)&0.001619(37)&4.31(10)\\
0.006&0.006&0.004&364.6(8)&739(2)&0.001668(31)&4.076(76)\\
0.006&0.006&0.008&368.4(7)&746(2)&0.001623(26)&4.023(64)\\
0.006&0.008&0.004&389.9(8)&739(2)&0.001673(31)&4.228(79)\\
0.006&0.008&0.008&393.7(7)&746(2)&0.001633(26)&4.182(66)\\
0.006&0.025&0.004&562.5(9)&739(2)&0.001767(37)&5.37(11)\\
0.006&0.025&0.008&566.1(8)&746(2)&0.001746(31)&5.362(95)\\
0.006&0.03&0.004&604.6(9)&739(2)&0.001798(39)&5.66(12)\\
0.006&0.03&0.008&608.3(9)&746(2)&0.001780(33)&5.67(11)\\
0.008&0.008&0.004&413.6(8)&837(2)&0.001863(32)&5.492(95)\\
0.008&0.008&0.008&417.4(7)&843(2)&0.001815(26)&5.414(80)\\
0.008&0.025&0.004&579.4(8)&837(2)&0.001937(35)&6.76(13)\\
0.008&0.025&0.008&582.9(8)&843(2)&0.001900(30)&6.70(11)\\
0.008&0.03&0.004&620.5(8)&837(2)&0.001964(37)&7.09(14)\\
0.008&0.03&0.008&624.0(8)&843(2)&0.001928(32)&7.03(12)\\
0.025&0.025&0.004&710.2(6)&1427(1)&0.003172(48)&20.89(32)\\
0.025&0.025&0.008&712.6(7)&1431(2)&0.003101(42)&20.51(28)\\
0.025&0.03&0.004&745.1(6)&1427(1)&0.003180(48)&21.45(33)\\
0.025&0.03&0.008&747.3(7)&1431(2)&0.003115(43)&21.09(29)\\
0.03&0.03&0.004&778.7(6)&1564(1)&0.003509(53)&26.52(40)\\
0.03&0.03&0.008&780.6(7)&1567(8)&0.00344(13)&26.1(1.1)\\
\hline
\end{tabular}
\end{center}
\normalsize
\end{table}

\begin{figure}[htp]
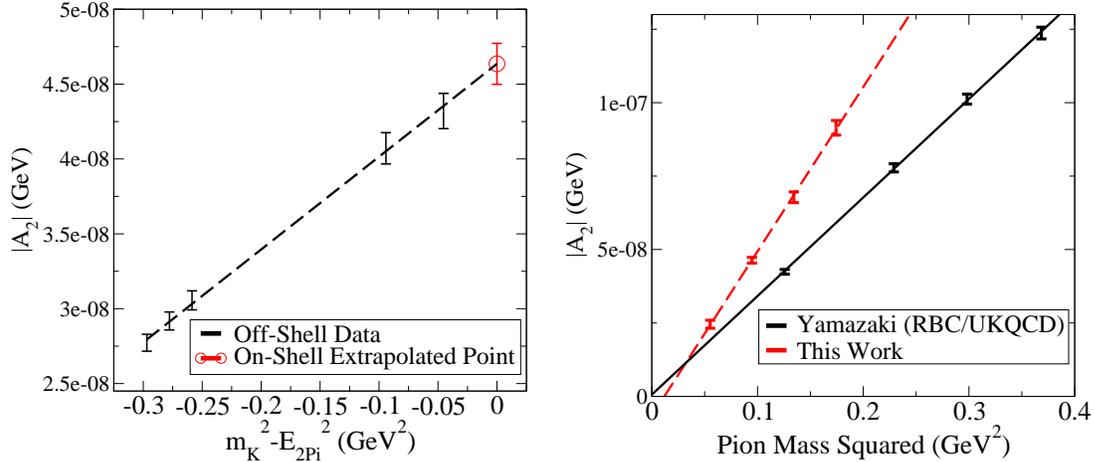

\centering
\includegraphics{on_shell_extrap}
\hspace{0.1 in}
\includegraphics{compare_w_takeshi}
\hfill
\vspace{-0.1 in}
\caption{\underline{Left:} Plot of the physical quantity $|A_2|$ vs. $m_K^2-E_{\pi\pi}^2$ for the fixed pion mass $m_\pi=308\text{ MeV}$, and fixed $E_{\pi\pi}=626\text{ MeV}$.  This plot is fit with a straight line, and extrapolated to the energy conserving point $m_K^2-E_{\pi\pi}^2=0$ (red circle). 
\underline{Right:} Plot of $|A_2|$ vs. $m_\pi^2$.  The preliminary data points of this work (red, dashed line) are shown alongside those from \cite{Takeshi} (black, solid line).  For the two largest pion masses in the present work the extrapolations to energy conserving kinematics are very large.  Thus the error bars for these masses are the systematic rather than statistical error, which is estimated as the difference between the extrapolated value obtained from a linear and from a quadratic fit.}\label{fig:A2}
\end{figure}

Since some of the mass combinations come close to conserving energy, it is not unreasonable to attempt to interpolate/extrapolate to energy conserving kinematics.  Specifically, holding the pion mass constant we plot $|A_2|$ as a function of $m_K^2-E_{\pi\pi}^2$ ($m_K$ is varied, $E_{\pi\pi}$ is constant), fit to a straight line, and extrapolate or interpolate to $m_K^2-E_{\pi\pi}^2=0$.  We do this for four different pion masses, and the plot plus extrapolation for $m_\pi=308\text{ MeV}$ is shown in Figure \ref{fig:A2} on the left.  

We now have values of $|A_2|$ for several pion masses extrapolated to kinematics which are energy conserving, as tabulated in Table 2.  In Figure \ref{fig:A2} on the right, we add the preliminary data in this work to the plot of $|A_2|\approx \text{Re}(A_2)$ vs. $m_\pi^2$ found in \cite{Takeshi}.

In Figure \ref{fig:A2} we see that the data in the present work does not appear to agree with that in \cite{Takeshi}.  This may be due to the fact that dynamical quarks are used in the present work whereas \cite{Takeshi} was done in the quenched approximation.  However, it could easily be due to the different methods used to set the lattice spacing (the $\rho$ mass for the lattices used in \cite{Takeshi} vs. the $\Omega$ mass for the lattices used in this work, see \cite{Enno_proceedings}), especially considering that $|A_2|$ is a dimensioned quantity that is proportional to $a^{-3}$ making it extremely sensitive to errors in $a$.  Furthermore, the lattice spacings in the two works are quite different ($a^{-1}=2.42\text{ GeV}$ in this work compared to $a^{-1}=1.31\text{ GeV}$ in \cite{Takeshi}) so that finite lattice spacing effects could come into play.  It has been checked that the same conventions for the normalization of $|A_2|$ were used in this work and in \cite{Takeshi}.

\begin{table}[htp]
\label{tb:A2_vs_mpi}
\caption{$|A_2|$ vs. $m_\pi^2$ after an interpolation/extrapolation to energy conserving kinematics.}
\begin{center}
\begin{tabular}{|c|c|}
\hline
$m_\pi^2$ ($\text{GeV}^2$)&$|A_2|$ ($10^{-8}$ GeV)\\
\hline
0.05521(68)&2.46(14)\\
0.09470(52)&4.64(14)\\
0.13432(40)&6.96(12)\\
0.17425(61)&9.50(20)\\
\hline
\end{tabular}
\end{center}
\end{table}

\vspace{-0.3 in}
\section{Conclusion}
\vspace{-0.15 in}
We have calculated the contribution of the operator $Q_{(27,1)}^{\Delta I=3/2}$ to $|A_2|$ in $\Delta I=3/2$ $K\to\pi\pi$ decays with 0 momentum pions on a $32^3\times 64$, $L_s=16$ 2+1 flavor DWF lattice.  We did a linear extrapolation to a unitary light sea quark mass from the two light sea quark masses of 0.004 and 0.008 studied in our calculation, but did not attempt to make the sea and valence strange quark masses agree.  We interpolated to energy conserving kinematics as well.  When this preliminary data is plotted alongside the data of \cite{Takeshi} in a graph of $|A_2|$ vs. $m_\pi^2$ we see a disagreement which could be explained by different methods of determining the lattice spacing and finite lattice spacing errors.  Also, the present work has the advantage of including 2+1 flavors of dynamical quarks, whereas all previous work has been in the quenched approximation.  In the present work there is only data for decays to zero momentum pions.

Future plans include doing the full $\Delta I=3/2$ calculation on $32^3\times 64$ lattices, but with significantly stronger coupling so that the resulting larger physical volume will permit using pion masses much closer to the physical value.  Momentum will be given to the pions using twisted boundary conditions \cite{Changhoan_thesis,Changhoan1,Changhoan2,Sachrajda_Villadoro} and the kinematics will be very close to physical kinematics.

{\bf Acknowledgements:}  I thank all of my colleagues in the RBC and UKQCD collaborations for helpful discussions and the development and support of the QCDOC hardware and software infrastructure which was essential to this work.  In addition I acknowledge Columbia University, RIKEN, BNL and the U.S. DOE for providing the facilities on which this work was performed.  This work was supported in part by U.S. DOE grant number DE-FG02-92ER40699.


\vspace{-0.15 in}
\begin{thebibliography}{99}
\vspace{-0.1 in}
  \bibitem{CPPACS} CP-PACS Collaboration, J.I. Noaki et al., \emph{Phys. Rev. D} {\bf 68} (2003) 014501 [{\tt hep-lat/0108013}]. 
  \bibitem{RBC} RBC Collaboration, T. Blum et al., \emph{Phys. Rev. D} {\bf 68} (2003) 114506 [{\tt hep-lat/0110075}].
  \bibitem{Changhoan_thesis} Changhoan Kim, Ph.D. Thesis, Columbia University, 2004.
  \bibitem{Changhoan1} Changhoan Kim and Christ, Norman H., \emph{Nucl. Phys. Proc. Suppl.} {\bf 119} (2003) 365 [{\tt hep-lat/0210003}].
  \bibitem{Changhoan2} C.H. Kim, \emph{Nucl. Phys. Proc. Suppl.} {\bf 140} (2005) 381.
  \bibitem{Takeshi} RBC and UKQCD Collaboration, T. Yamazaki, [{\tt hep-lat/0807.3130}].
  \bibitem{Ciuchini} M.~Ciuchini et al., \emph{Z. Phys. C} {\bf 68} (1995) 239 [{\tt hep-ph/9501265}]. 
  \bibitem{Buchalla} G.~Buchalla et al., \emph{Rev. Mod. Phys.} {\bf 68} (1996) 1125 [{\tt hep-ph/9512380}].
  \bibitem{Enno_proceedings} Enno E. Scholz, These proceedings.
  \bibitem{Sam_thesis} Shu Li, Ph.D. Thesis, Columbia University, 2008.
  \bibitem{CKelly} Chris Kelly, These proceedings.
  \bibitem{Lellouch_Luscher} L. Lellouch and Luscher, M., \emph{Commun. Math. Phys.} {\bf 219} (2001) 31 [{\tt hep-lat/0003023}].
  \bibitem{Sachrajda_Villadoro} C.T. Sachrajda and Villadoro, G., \emph{Phys. Lett. B} {\bf 609} (2005) 73 [{\tt hep-lat/0411033}]

\end{thebibliography}
\end{document}